\documentclass[aps,twocolumn,nofootinbib,longbibliography]{revtex4-1}
\usepackage[babel]{csquotes}
\usepackage{graphicx}
\usepackage{amsmath,amssymb}
\usepackage[colorlinks,citecolor=blue,linkcolor=blue,urlcolor=blue]{hyperref}
\usepackage{mathrsfs}
\usepackage{enumerate}
\def\be{\begin{equation}}
\def\ee{\end{equation}}
\def\H{{\cal H}}
\def\tr{{\rm tr}}

\begin{document}

\title{Why do we remember the past and not the future? \\ 
The `time oriented coarse graining' hypothesis}

\author{Carlo Rovelli}

\affiliation{Aix Marseille Universit\'e, CNRS, CPT, UMR 7332, 13288 Marseille, France\\
Universit\'e de Toulon, CNRS, CPT, UMR 7332, 83957 La Garde, France.}

\date{\small\today}

\begin{abstract}
\noindent 
Phenomenological arrows of time can be traced to a past low-entropy state.  Does this imply the universe was in an improbable state in the past?  I suggest a different possibility:  past low-entropy depends on the coarse-graining implicit in our definition of entropy. This, in turn depends on our physical coupling to the rest of the world.  I conjecture that \emph{any} generic motion of a sufficiently rich system satisfies the second law of thermodynamics, in either direction of time, for \emph{some} choice of macroscopic observables.  The low entropy of the past could then be due to the way we couple to the universe (a way needed for us doing what we do), hence to our natural macroscopic variables, rather than to a strange past microstate of the world at large.

\end{abstract}

\maketitle
    
\section{Introduction}

The second law of thermodynamics states that entropy does not decrease in time.  This is not in contradiction with the time  (more precisely, CPT) reversal invariance of the microscopic laws, because we  use the second law to study the evolution of states with \emph{initial} low entropy.  Fixing the \emph{initial} conditions is what breaks time reversal symmetry. But why do we use \emph{initial} conditions, if microphysics is time reversal invariant?    

The reason is that the world we inhabit is strongly time-asymmetric: observed phenomena display marked and consistent arrows of time.  The discussion about the interpretation of this fact has been extensive, and I am not going to recall it here; the conclusion is that there is substantial consensus that all arrows of time can be traced to the fact (or repressed as the statement) that the world at large had low entropy in the past (see for instance \cite{Lebowitz1993}). We live in a universe with  continuous abundant entropy production.  Here I take this conclusion for granted and consider a further question: why was the entropy of the world low in the past?

Is it because the universe happened to be born in a very peculiar microstate?  (For a recent discussion in the context of cosmology see for instance \cite{Carroll2010}.) Is it because of some not yet known fundamental law, for instance Roger Penrose's hypothesis that initial singularities (but not final ones) have vanishing Weil curvature \cite{Penrose:1979fk}? 

Here I suggest a different possibility.  The entropy of a system depends on the determination of a set of macroscopic observables $A_n$, namely on a ``coarse-graining" of the microscopic state space of the system. I suggest that the past entropy of the world was not low because of its microstate being peculiar: rather, it is the coarse graining we use to describe it that makes its entropy low. 

More precisely, I consider the following: 
\begin{description} 
\item[Conjecture]  Any generic microscopic motion of a sufficiently rich system satisfies the second law (in either time direction)  for a suitable choice of macroscopic observables.  
\end{description}
Below I give some argument supporting the plausibility of this conjecture.  If the conjecture holds, then we have a new way for facing the puzzle of the arrow of time:  the universe was not in any special state in the past; rather, the universe is sufficiently rich to admit a description in terms of coarse grained observables with respect to which entropy increases.  

Now, the choice of coarse graining in the thermodynamical description of a system is not free: it depends on the system's \emph{couplings}.  A system that couples to the rest of the universe via macroscopic variables for which entropy was low in the past interacts with a universe which is strongly time oriented. As we do.  Such a coupling might therefore characterize subsystems where information gathering  creatures such as ourselves can exist. 

Entropy increase, and therefore all phenomena related to time flow, might follow not from a peculiarity of the microstate of the universe, but from the peculiarities of our coupling to the rest of the universe.

\section{The conjecture}

I start by making the formulation of the conjecture a bit more precise.  I introduce it using classical mechanics, namely disregarding quantum theory.  I extend it to the quantum theory in the last section. Quantum theory, I think, reinforces the plausibility of the idea, but the classical picture is more clear.  In this context it is convenient to use Gibbs' formulation of statistical mechanics rather than Boltzmann's, because the Boltzmann's formulation takes for granted the split of a system in a large number of equal subsystems (the individual particles), and this may not be the right way of thinking in quantum field theory, where the story becomes in fact more interesting. 

Consider a classical system with many degrees of freedom, in a (``microscopic") state $s$, element of a phase space $\Gamma$, evolving in time as $s(t)$.  Let $A_n$, be a set of (``macroscopic") observables --real functions on $\Gamma$--,  labeled by the index $n$.  Given $s$ and $A_n$,  (micro canonical) entropy can be  defined as the logarithm of the number of states that have the same $A_n$'s values as $s$.  If phase space is continuous, with a (time invariant) measure $ds$, entropy is then 
\be
 S_{s,A_n}=\log \int_\Gamma ds' \prod_n \delta(A_n(s')-A_n(s)).
\ee
Notice that this definition applies to any microstate.\footnote{This equation defines entropy up to an an additive factor, because phase space volume has the dimension of $[Action]^N$, where $N$ is the number of degrees of freedom. This incompleteness is settled by quantum theory, which introduces the natural unit of action, the Planck constant, whose physical meaning is to determine the minimum empirically distinguishable phase space volume, namely the maximal amount of information in a state. See \cite{Haggard:2013fx}.}

  As the microstate $s$ evolves in time so does entropy
\be
S_{s,A_n}(t)\equiv S_{s(t),A_n}.
\ee  
The conjecture states that if the system is sufficiently complex and ergodic, for most paths $s(t)$ that satisfy the dynamics and for each choice of a direction in $t$, there is a family of observables $A_n$ such that 
\be
\frac{dS_{s,A_n}}{dt}\ge 0. 
\ee  
In other words, \emph{any} motion appears to have initial low entropy (and then non decreasing entropy) under \emph{some} coarse graining. 

\section{Balls in a box}

A simple example illustrates the conjecture.   Consider a set $\Sigma$ of $N$ distinguishable balls that move in a box, governed by a time reversible and sufficiently ergodic dynamics. Let the box have an extension $x\in[-1,1]$ in the direction of the $x$ coordinate, and let it be ideally divided in two halves by $x=0$.  For any subset of balls $\sigma\subset \Sigma$ define the observable $A_\sigma$ to be the mean value of the $x$ coordinate of the balls in $\sigma$.  That is, if $x_b$ is the $x$ coordinate of the ball $b$, define
\be
    A_\sigma = \frac{\sum_{b\in\sigma} x_b}{\sum_{b\in\sigma}1}.
\ee

Let $s(t)$ be a generic actual motion of the system, say going from $t=t_i$ to $t=t_f$.  Let $\sigma_i$ be the set of the balls that are at the left of $x=0$ at $t=t_i$.  The macroscopic observable $A_i\equiv A_{\sigma_i}$ defines an entropy that in the large $N$ limit and  for most motions $s(t)$ satisfies
\be
\frac{S_{s,A_i}(t)}{dt}\ge 0. 
\ee
This is the second law of thermodynamics.  Notice that it holds for a \emph{generic} motion $s(t)$. 

Given the \emph{same} motion $s(t)$, we can define a different observable as follows. Let $\sigma_f$ be the set of the balls that are at the left of $x=0$ at $t=t_f$.  The macroscopic observable $A_f \equiv  A_{\sigma_f}$ defines an entropy that satisfies
\be
\frac{S_{s,A_f}(t)}{dt}\le 0. 
\ee
This is again second law of thermodynamics, but now in the reversed time $-t$.  It holds for the same (generic) motion $s(t)$, but a different observable family. 

This is pretty obvious on physical grounds: if at initial time we ideally color in red all the balls at the left of $x=0$ then the initial state is low entropy with the respect to the coarse graining given by color, and most motions are going to mix the balls and raise the entropy.  

The point is simple: for any motion there is a macroscopic family of observables with respect to which the state at a chosen end of the motion is low entropy.   I call these observables, ``time oriented observables".
These observable are easily determined by the state itself. 

\section{Observables are interactions}

The fact that thermodynamics and statistical mechanics require a coarse graining, namely a ``choice" of macroscopic observables, appears at first sight to introduce a curious element of subjectivity into physics,  clashing with the objectivity of the predictions of science. 

But of course there is nothing subjective in thermodynamics. A cup of hot tea does not cool down because of what I know or do not know about its molecules. The ``choice" of macroscopic observables is dictated  by the ways the system under consideration couples.  The macroscopic observables of the system are those coupled to the exterior (in thermodynamics, those that can be manipulated and  macroscopically measured). The thermodynamics and the statistical mechanics of a system defined by a set of macroscopic observables $A_n$ describes (objectively) the way the system interacts when coupled to another system via these observables.

For instance, the behavior of a box full of air is going to be described by a certain entropy function if the air is interacting with the exterior via a piston that changes its volume $V$.  But the \emph{same} air is going to be described by a \emph{different} entropy function if it interacts with the exterior via \emph{two} pistons with  filters  permeable to Oxygen and Nitrogen respectively. In this case, the macroscopic observables are others and a chemical potential enters the game.   It is not our abstract ``knowledge" about the relative abundance of Oxygen and Nitrogen that matters: it is the presence or not of a  physical coupling of this quantity to the exterior to determine which entropy describes the phenomena. Different statistics and thermodynamics of the same box of air describe different interactions of the box with the exterior. 

In light of this consideration, let us reconsider the box of the previous section replacing the abstract notion of ``observable" by a concrete interaction between subsystems. This is done in the next Section.

\section{Time-oriented subsystems}

Say we have the $N$ balls in a box as above, but now we add a new set of $2^N$ ``small" balls\footnote{$2^N$ is the number of subsets of $\Sigma$, namely the cardinality of its power set.}, with negligible mass, that do not interact among themselves but interact with the previous (``large") balls as follows. Each small ball is labeled by a subset $\sigma\subset\Sigma$ and is attracted by the balls in $\sigma$ and only these, via a force law such that the total attraction is in the direction of the center of mass $A_\sigma$ of the balls in $\sigma$. 

Thus, generically a small ball interacts with a large number of large balls, but it does so only via a single variable:  $A_\sigma$. Therefore it interacts with a thermodynamical system, for which $A_\sigma$ is the macroscopic observable. For each small ball $\sigma$, the ``rest of the universe"  behaves as a thermal system with entropy $S_{s,A_\sigma}$.  

As shown above, given a \emph{generic} motion $s(t)$ there will generically be at least one small ball,  the ball $\sigma_i$ for which the entropy of the rest of the box is never decreasing in $t$ (in the thermodynamical limit of large $N$).

To make the example more intuitively compelling, imagine now that the box is the universe and each ``small" ball $\sigma$ is itself a large system with very many degrees of freedom.  Then generically there is at least one of these, namely $\sigma_i$ (but in fact many) for which the rest of the universe is seen as having a low-entropy initial state. 

Since $\sigma_i$ interacts thermodynamically with a universe which was in a low-entropy state in the past, the world seen by $\sigma_i$ appears organized in time: observed phenomena display marked and consistent arrows of time, which single out one direction.  $\sigma_i$ interacts with a world where entropy increases, hence ``time flows" in one specific direction. The world seen by $\sigma_i$ may include dissipation, memory, traces of the past, and all the many phenomena that characterize the universe as we see it.  Within such subsystem $\sigma_i$ time oriented phenomena that require the growth of entropy, such as evolution or accumulation of knowledge, can take place. 

I call such a subsystem a ``time oriented subsystem".

Could this picture be related to the reason why we see the universe as a system with a low-entropy initial state?  Could the low entropy of the universe characterize our own coupling with the universe, rather than a peculiarity of the microstate of the universe?

\section{Structure formation}

If the hypothesis put forward in the previous section hold, we may ask how come we happen to be part of a special subsystem that couples in a  peculiar manner.   There is a natural answer: because this is the precondition for us to be what we are, in a wide sense.  We live in time, our existence depends strongly on entropy production. Innumerable aspects of our own existence depends on the fact of being in a situation of local strong entropy production: biological evolution, biochemistry, life itself, memory, knowledge acquisition, culture...  Being part of a time oriented subsystem could simply be the condition for all this? 

A (loose) analogy: we live in a region of the universe with liquid water.  Would it make sense to ask for a reason why we are in such an unusual place?  The answer is clearly no: we are that sort of things that develop where liquid water is.   So, our inhabiting theses quarters of the universe is no more strange than me being born in a place where people happen to speak my own language. 

Consider the thermodynamical framework of life on Earth.  There is a constant flow of electromagnetic energy on Earth: incoming radiation from the Sun and outgoing radiation towards the sky. Microscopically, this is a certain complicated solution of Maxwell equation, whatever this is.  But as far as life is considered, most details of this solution  (such as the precise phase of a single solar photon falling on the Pacific ocean) are irrelevant.

What matters are energy and frequency integrated over small regions. Now the incoming energy is the same as the outgoing energy, but not so for the frequency.  The Earth receives energy  $E$ (from the Sun) at higher frequency $\nu_i$ and emits energy (towards the sky) at lower frequency $\nu_f$.   This is a fact about the actual solution of the Maxwell equations in which we happen to live.  If we take energy and frequency as macroscopical observables, then entropy is defined.  Roughly, entropy counts the number of photons; at frequency $\nu$ the number of photons in a wave of energy $E$ is $N=E/\hbar\nu$. If the received energy is emitted at lower frequency, the emitted entropy $S_f$ is higher than the received entropy $S_i$.  The process produces entropy: $S_f\gg S_i$.  This entropy production is not a feature of the  solution of the Maxwell equations alone:  it is a feature of this solution \emph{and} a set of macroscopic observables (energy and frequency) to which living systems couple.  

Any system on Earth whose dynamics is governed by interactions with $E$ and $\nu$ has a source of negative entropy at its disposal.  This is what is exploited by life on Earth to build structure and organization.

Could it be that entropy growth is not a property of the universe by itself, but rather a property of the particular coupling to the rest of the universe that allows us to be what we are? 

\section{The subject, the object and environment}

Any description of the  world is based on the availability of relevant information about a portion of this world.  Having information about a part of the world implies being correlated with that part. Thus information depends on the existence of a \emph{physical} correlation between two systems. For instance I have information on the color of the sky if there is a certain correlation between the state of the neurons in my brain and the color of the sky.  This correlation is the information I have about the sky. An apparatus has information about a variable of the system because after the measurement the  pointer position is correlated with the value of the variable. A phone book has information because the position of its ink's drop are correlated to the memories of the phone company's switches. 

The physical notion underlying information is therefore Shannon's relative information 
\be
S_{12}=\log[N_1\times N_2]-\log N_{12},
\ee
which measures the constraint reducing the states available to a system to a number $N_{12}$ smaller than the number of states $N_1\times N_2$ in the Cartesian product of the state spaces of its individual subsystems. 

Information is the fact that certain product states (sky blue and my neurons having an imprint of grey sky, or sky grey and my neurons having an imprint of blue sky) are impossible or improbable.

These correlations are established by physical interactions.  For instance, I get information about the weather by opening the window and interacting with the photons coming from the outside. This interaction determines a correlation between by neurons and the sky and therefore establishes information.  (I have discussed these points more in detail in \cite{Rovelli2013f}.)

Correlations vary because of dynamical interactions between the two systems and the rest of the world.  The proper setting for statistical considerations is therefore formed by \emph{three} (and not two) components, which can be called (i) the subject, (ii) the object and (iii) the environment.  

The state of the object is described by the macroscopic variables.  The subject is the part of the world which interacts with the object and gets correlated with it, thus holding information (in the sense of Shannon) about these variables.  The environment is the ensemble of the other variables the object interacts with.  

Notice that interactions between the object and the environment \emph{raise} entropy, because part of the information held by the subject may become irrelevant.  While interactions between the subject and the object \emph{lower} the entropy, because new information becomes available. This triple part structure underlying statistical mechanics and this double mode of evolution of entropy have been emphasized in particular by Max Tegmark \cite{Tegmark2012a}.

Past low entropy, therefore, should not be regarded as a feature of the microstate of the world, but rather as something pertaining a peculiar split of the world.  

\section{Quantum theory}

Quantum phenomena generate entropy distinct from the classical one generated by coarse graining: entanglement entropy. The state space of any quantum system is described by a Hilbert space $\H$, with a  linear structure that plays a major role for physics.  If the system can be split into two components, its state space splits into the tensor product of two Hilbert spaces: $\H=\H_1\otimes\H_2$, each carrying the corresponding subset of observables. Because of the linearity, a generic state is not a tensor product of component states; that is, in general $\psi\ne\psi_1\otimes\psi_2$. This is entanglement. Restricting the observables to those of a subsystem, say system $1$, determines a quantum entropy over and above classical statistical entropy. This is measured by the von Neumann entropy $S=-\tr[\rho\log\rho]$ of the density matrix $\rho=\tr_{\H_2}|\psi\rangle\langle\psi|$. 

The conjecture above can be adapted to this case. In any sufficiently rich quantum system (this means: with a sufficient complex algebra of observables: remember all separable Hilbert space are isomorphic), given a generic state evolving in time as $\psi(t)$, there is a split of the system into subsystems such that the von Neumann entropy is low at initial time and increases in time.  

The point here is not to assume the tensorial structure of $\H$ a priori. Instead, given a generic state, we can find a tensorial split of $\H$ which sees von Neumann entropy grow in time. 

This is easy to prove.  A separable Hilbert space admits many discrete bases $|n\rangle$. Given any $\psi\in\H$, we can always choose a basis $|n\rangle$ where $\psi=|1\rangle$. Then we can consider two Hilbert space,  $\H_1$ and $\H_2$, with bases $|k\rangle$ and $|m\rangle$, and map their tensor product to $\H$ by identifying  $|k\rangle \otimes |m\rangle$ with the state $|n\rangle$ where $(k,m)$ appear in the $n$-th position of the Cantor ordering of the $(n,m)$ couples  ((1,1),(1,2),(2,1),(1,3),(2,2),(3,1),(1,4)...).  Then, $\psi= |1\rangle \otimes |1\rangle$ is a tensor state and has vanishing von Neumann entropy. 

Therefore for any time evolution $\psi(t)$ there is a split of the system into subsystem such that the initial state has zero entropy. Then, growing and decreasing of (entanglement) entropy is an issue about how the universe in split into subsystems, not a feature of the overall state of things (on this, see again \cite{Tegmark2012a}).  

Entropic peculiarities of the past state of the universe should not be searched in the cosmos at large. They should be searched in the spilt, and therefore the macroscopic observables, relevant for us.   Time asymmetry, and therefore ``time flow", might be a feature of a subsystem to which we belong,  features needed for information gathering creatures like us to exist, and not features of the universe at large.

\centerline{------}  

I am indebted to Hal Haggard and Angelo Vulpiani for discussions and suggestions.


\end{document}